\documentclass[manuscript]{aastex6}

\usepackage{graphicx,fancyhdr,rotating,amsmath,epsf,txfonts,epstopdf,multirow,psfig}
\usepackage{graphicx,graphics,xcolor,harpoon}
\usepackage{natbib}

\begin{document}
\title{Magnetic braids in eruptions of a spiral structure in the solar atmosphere}
\author{
Zhenghua Huang\altaffilmark{1}, Lidong Xia\altaffilmark{1}, Chris~J.~Nelson\altaffilmark{2,3}, Jiajia Liu\altaffilmark{2,4}, Thomas Wiegelmann\altaffilmark{5}, Hui Tian\altaffilmark{6}, James~A.~Klimchuk\altaffilmark{7},Yao Chen\altaffilmark{1}, Bo Li\altaffilmark{1}
}
\altaffiltext{1} {Shandong Provincial Key Laboratory of Optical Astronomy and Solar-Terrestrial Environment, Institute of Space Sciences, Shandong University, Weihai 264209, Shandong, China}
\altaffiltext{2}{School of Mathematics and Statistics, Hicks Building, University of Sheffield, Hounsfield Road, Sheffield S3 7RH, UK}
\altaffiltext{3}{Astrophysics Research Centre, School of Mathematics and Physics, Queen's University, Belfast BT7 1NN, Northern Ireland, UK}
\altaffiltext{4}{CAS Key Laboratory of Geospace Environment, Department of Geophysics and Planetary Sciences, University of Science and Technology of China, Hefei 230026, China}
\altaffiltext{5}{Max-Planck-Institut f\"ur Sonnensystemforschung, Justus-von-Liebig-Weg 3, D-37077 G\"ottingen, Germany}
\altaffiltext{6}{School of Earth and Space Sciences, Peking University, Beijing 100871, China}
\altaffiltext{7}{NASA Goddard Space Flight Center, Code 671, Greenbelt, MD 20771, USA}

\begin{abstract}
We report on high-resolution imaging and spectral observations of eruptions of a spiral structure in the transition region,
which were taken with the Interface Region Imaging Spectrometer (IRIS), the Atmospheric Imaging Assembly (AIA) and the Helioseismic and Magnetic Imager (HMI).
The eruption coincided with the appearance of two series of jets, with velocities comparable to the Alfv\'en speeds in their footpoints.
Several pieces of evidence of magnetic braiding in the eruption are revealed, including localized bright knots, multiple well-separated jet threads, transition region explosive events and the fact that all these three are falling into the same locations within the eruptive structures.
Through analysis of the extrapolated three-dimensional magnetic field in the region, we found that the eruptive spiral structure corresponded well to locations of twisted magnetic flux tubes with varying curl values along their lengths. The eruption occurred where strong parallel currents, high squashing factors, and large twist numbers were obtained.
The electron number density of the eruptive structure is found to be $\sim3\times10^{12}$\,cm$^{-3}$, indicating that significant amount of mass could be pumped into the corona by the jets.
Following the eruption, the extrapolations revealed a set of seemingly relaxed loops, which were visible in the AIA\,94\,\AA\ channel indicating temperatures of around 6.3 MK.
With these observations, we suggest that magnetic braiding could be part of the mechanisms explaining the formation of solar eruption and the mass and energy supplement to the corona.
\end{abstract}
\maketitle

\section{Introduction}
Several major problems remain puzzling in the field of solar physics, including how the solar magnetic field drives large-scale eruptive phenomena, how the solar wind is generated and accelerated, and how the solar corona is heated to millions of degrees of Kelvin\,\citep{Shibata2011lrsp,Klimchuk2006SoPh,Tu2005Sci}.
Magnetic reconnection, the physical process whereby local magnetic fields are rearranged, converting magnetic energy into kinetic and thermal energy, could be one of the keys to understand these problems\,\citep[e.g.][]{priest2000book,Chen2011lrsp,Shibata2011lrsp,Li2016NatPh,Sun2015NatCo,Xue2016NatCo,Fisk2003JGR, Tu2005Sci, Antiochos2011ApJ, Crooker2012JGRA, Su2013, Klimchuk2006SoPh, Priest1998Nature, Schrijver1998Nature}.
Proposed by \citet{Parker1983ApJa,Parker1983ApJb}, magnetic braids
occur whenever twisted magnetic flux tubes wind around one another,
and they are one of the preferred topologies that generate the magnetic reconnection required to power the corona\,\citep{Parker1983ApJa,Parker1983ApJb, Parker1988ApJ, Schrijver1998Nature, Schrijver2007ApJ}. 
A few pieces of evidence in favour of magnetic braiding have come from the high resolution intensity maps of the Sun\,\citep[i.e., Hi-C observations,][]{cirtain2013nature}. However, direct observational evidence remains sparse, because the braiding structures are small and quickly untied at their early stage through reconnection and the resulting heat can be quickly spreading along the magnetic strands, thus difficult to resolve in real observations\,\citep{WilmotSmith2009ApJ,Pontin2017ApJ}.

\par
Here, we report on high-resolution observations that show evidence of magnetic braids associated with an eruption of a structure in the solar transition region, with the heating of a bundle of loops and a bulk of plasma flow into the upper solar atmosphere.
Based on the present observations, we suggest that magnetic braiding could be part of the mechanisms explaining the three major problems in solar physics.

\section{Observations of the active event}
The data were collected by the Interface Region Imaging Spectrograph\,\citep[IRIS,][]{DPB2014IRIS} and both the Atmospheric Imaging Assembly\,\citep[AIA,][]{Lemen2012aia} and Helioseismic and Magnetic Imager\,\citep[HMI][]{Schou2012hmi} aboard the Solar Dynamics Observatory\,\citep[SDO;][]{Pesnell2012sdo} on 2014 June 11.
The IRIS observations include SJ images sampled in four passbands (1330\,\AA, 1400\,\AA, 2796\,\AA\ and 2832\,\AA) with a resolution of 0.35$^{''}$\ and cadences of 16\,s, 16\,s, 16\,s and 86\,s, respectively. In addition to this, a spectral slit scanned the event repeatedly with 96 steps, 0.35 arcsec step size, and 4 second exposure times to provide the spectral data of the region.

\par
\begin{figure*}
\includegraphics[clip,trim=0.5cm 0.3cm 2.5cm 0cm,width=\textwidth]{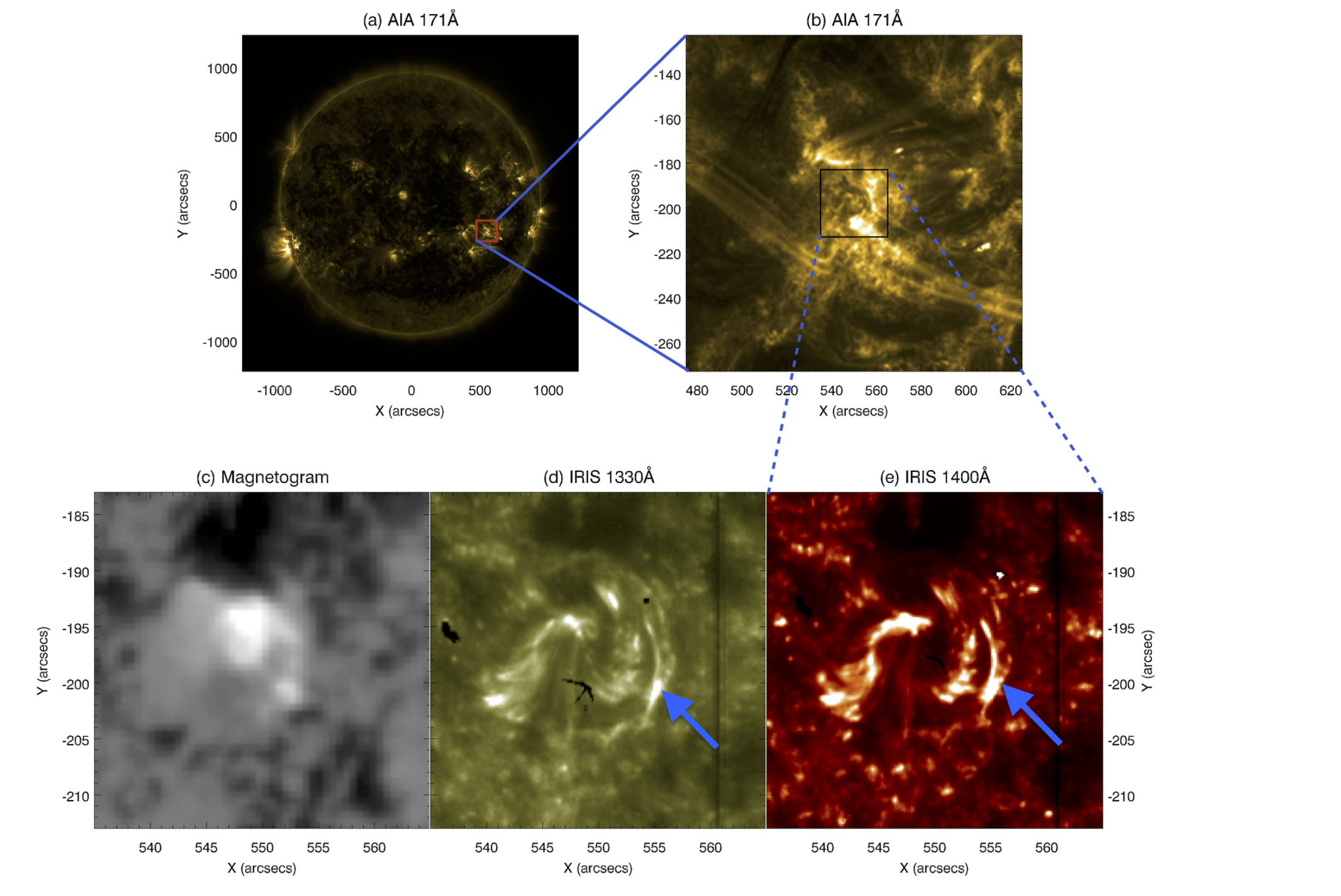}
\caption{
Context images displaying the spiraled filament analysed in this article. (a) A full disk image of the Sun on 2014 June 11 at 08:40 UT sampled by the AIA 171\,\AA\ filter (~630 000 K). (b) A zoomed-in view of the region of interest as denoted by red box in panel (a). The spiral structure is outlined by the black box. (c) The line-of-sight magnetogram co-spatial to the black box overlaid on panel (b). The image is saturated at $-$1\,000\,G (black) and 1\,000\,G (white). (d) The spiral structure seen in the IRIS 1330\,\AA\ passband (25\,000\,K). (e) The spiral structure seen in the IRIS 1400\,\AA\ passband (80\,000\,K). The arrows in panels (d) and (e) point to the ``spiral arm'' that will erupt. (An animation is given online.)
}
\label{figfov}
\end{figure*}
\begin{figure*}
\includegraphics[clip,trim=1cm 0.5cm 2.5cm 0cm,width=\textwidth]{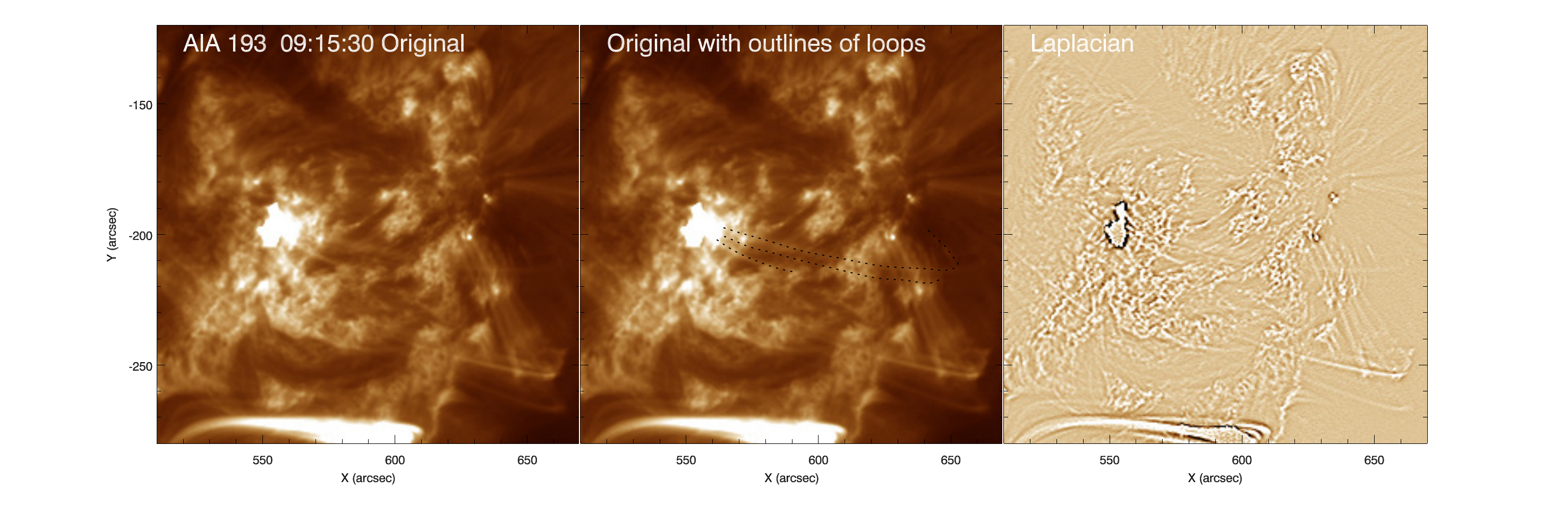}
\caption{The studied region with a zoomed-out view in the AIA 193\,\AA\ passband at 09:15:30\,UT. Left panel: the original image. 
Middle panel: original image with a few large coronal loops indicated by dotted lines, which are associated with the large field line system shown in Fig.\,\ref{figsk}, and the studied event is corresponding to their eastern footpoints (bright region around coordinates of [555,-195]). Right panel: The original image with the loops enhanced by applying a Laplacian sharpening filter.
}
\label{fig193_large}
\end{figure*}
\begin{figure*}
\includegraphics[clip,trim=1.5cm 0cm 0cm 0cm,width=\textwidth]{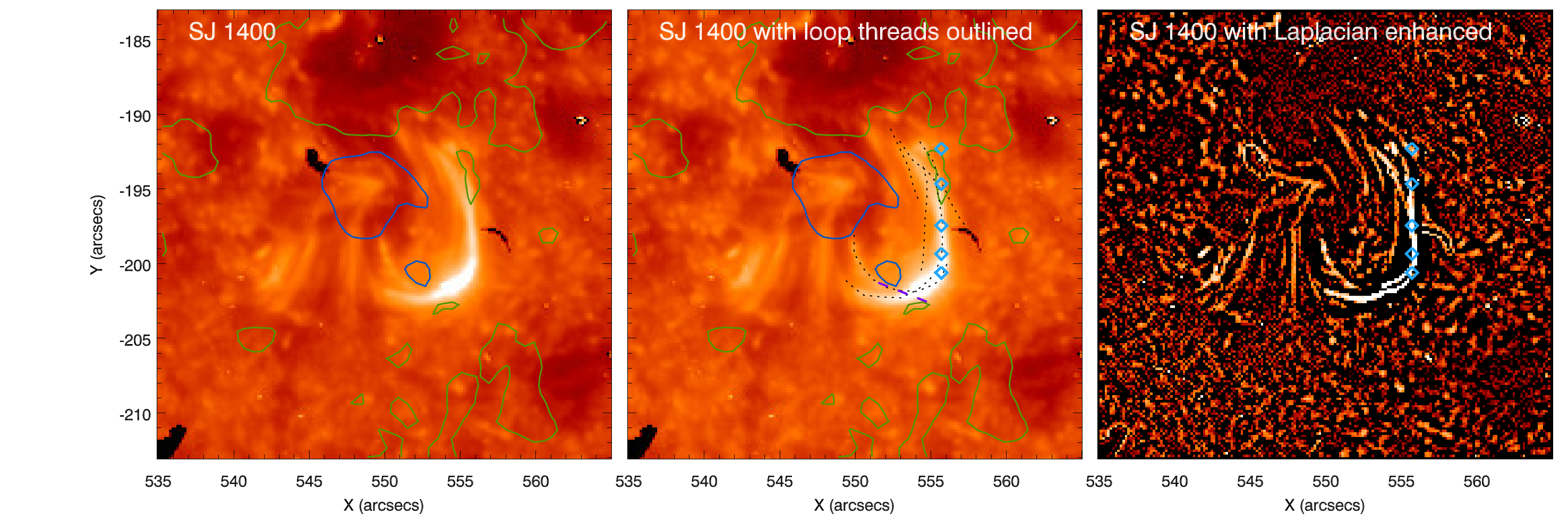}
\caption{The event seen in the IRIS SJ 1400\,\AA\ passband at 09:08:01\,UT, when the structure just began to erupt. Left: the event in the SJ 1400\,\AA\ passband presented with a logarithmic scaling. Middle: The same as the left panel but with visible strands in the eruptive structure outlined by black dotted lines. The over-lying loop system that crosses at the first reconnection point is outlined by the purple dashed line (determined from SJ 1400\,\AA\ image at 09:05\,UT). Right: The same image as the left panel, but enhanced by the Laplacian sharpening filter. The contours in the left and middle panels are representative of line-of-sight magnetic strength at levels of $-$500\,G (green) and 500\,G (blue). The cyan diamonds in the middle and right panels are the locations where the explosive events occurred.}
\label{figs1400}
\end{figure*}

The region of interest was part of an Active Region 12080 as shown in Figure\,\ref{figfov}. 
The line-of-sight photospheric magnetogram (Figure\,\ref{figfov}c) displays some kilo-Gauss positive (white) polarity magnetic fields surrounded by negative (black) polarity. 
IRIS data show a spiral structure in the 1330\,\AA\ ($\sim$25\,000\,K, see Figure\,\ref{figfov}d) and 1400\,\AA\ ($\sim$80\,000\,K, see Figure\,\ref{figfov}e) passbands (representative of the transition region). 
We suspect that such spiral structure could have formed via some swirling motion in the centroid, as similar as that have been observed in such as macrospicules\,\citep{2010A&A...510L...1K}, network magnetic fields\,\citep{2011ApJ...741L...7Z},
tornados and giant-tonardos\,\citep[e.g.][]{2012ApJ...752L..22L,2012ApJ...756L..41S,Wedemeyer2012nature,2013ApJ...774..123W} and swirls\,\citep[e.g.][]{Wedemeyer2009AA,DP2014Sci} in the solar atmosphere. To identify such swirling motion (if any), one should require high resolution chromospheric data (such as H$\alpha$) that are not available with the current data. However, its formation mechanism is not the scope of the present study.
The spiral structure is connected to a remote region by a large loop system that is clearly seen in the coronal images (see Figure\,\ref{fig193_large}).
The structure was actively evolving (see the animations 1--3), and one of its eruptions on June 11 2014 was recorded by the instruments above. A timeline of the eruption can be found in Table\,\ref{tab_evolution}.
The eruption occurred at one ``spiral arm'' of the structure, which manifests as a cluster of magnetic loops that are bundled together (see Figures\,\ref{figfov}\&\ref{figs1400} and the online animation).
This provides a favorable environment for magnetic braiding

\par
\begin{figure*}
\includegraphics[width=\textwidth]{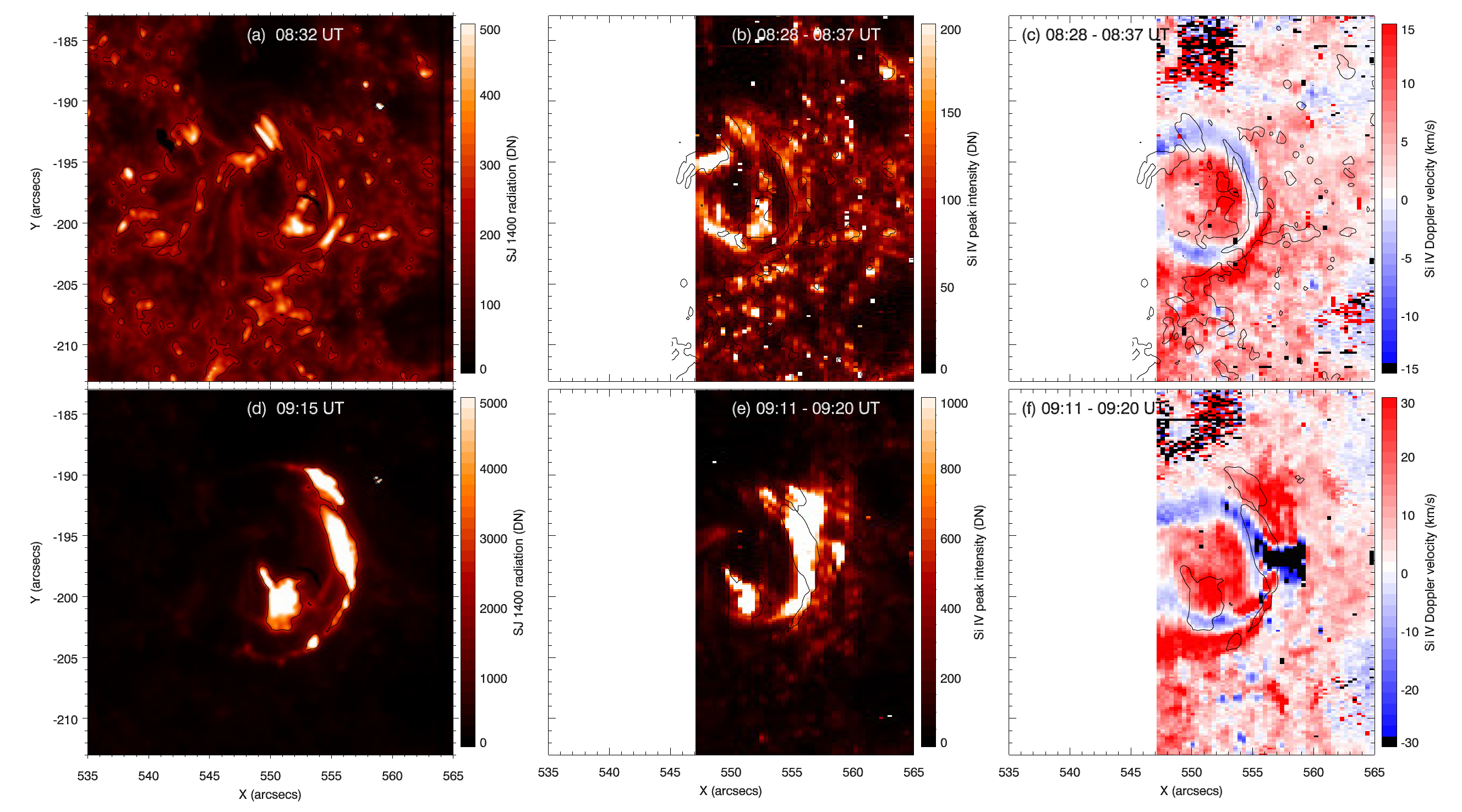}
\caption{Imaging and spectroscopy of the spiral structure before (top row) and during (bottom row) the eruption. (a) The event in the IRIS SJ 1400\,\AA\ channel at 08:32\,UT; (b) the region of the structure canned by the spectral slit between 08:28\,UT and 08:37\,UT using the spectral line of Si IV 1403\,\AA; (c) the corresponding Dopplergram of the region. Panels (d--f) show the same as (a--c), but for the event during its eruption around 09:15\,UT. The contours outline the structures of the eruptive events in the IRIS SJ 1400\,\AA\ images at the time shown in the panels on the left (a \& d).}
\label{figs1}
\end{figure*}

The AIA data reveal that the cluster of eruptive loops also consists of dark filament-like features, possibly containing cool materials (animation 1). The cool filament materials appear to be ejected during the eruption, in a manner similar to many eruptive phenomena on the Sun\,\citep{Sterling2015Nature, 2016ApJ...830...60H,2017ApJ...835...35H, Chen2011lrsp, Shibata2011lrsp}.
In order to derive the line-of-sight velocity (Doppler velocity) of the region, we fitted Si\,{\sc iv}\,1403\,\AA\ spectral profile with a Gaussian function.
The Doppler velocities were obtained by measuring the shift of the fitted profile from the average profile of the whole region (which was assumed to be at rest).
The Doppler velocities of the region are shown in Figure\,\ref{figs1}.
The velocity map of the pre-eruption stage indicates at least two intertwined loop systems with flows in opposite directions (Figure\,\ref{figs1}). 
As seen from the following analysis, the intersections of these loops correspond to one of the major sites of subsequent magnetic reconnection.

\par
\begin{figure*}
\includegraphics[clip,trim=1.5cm 0.3cm 2.8cm 0.5cm,width=\textwidth]{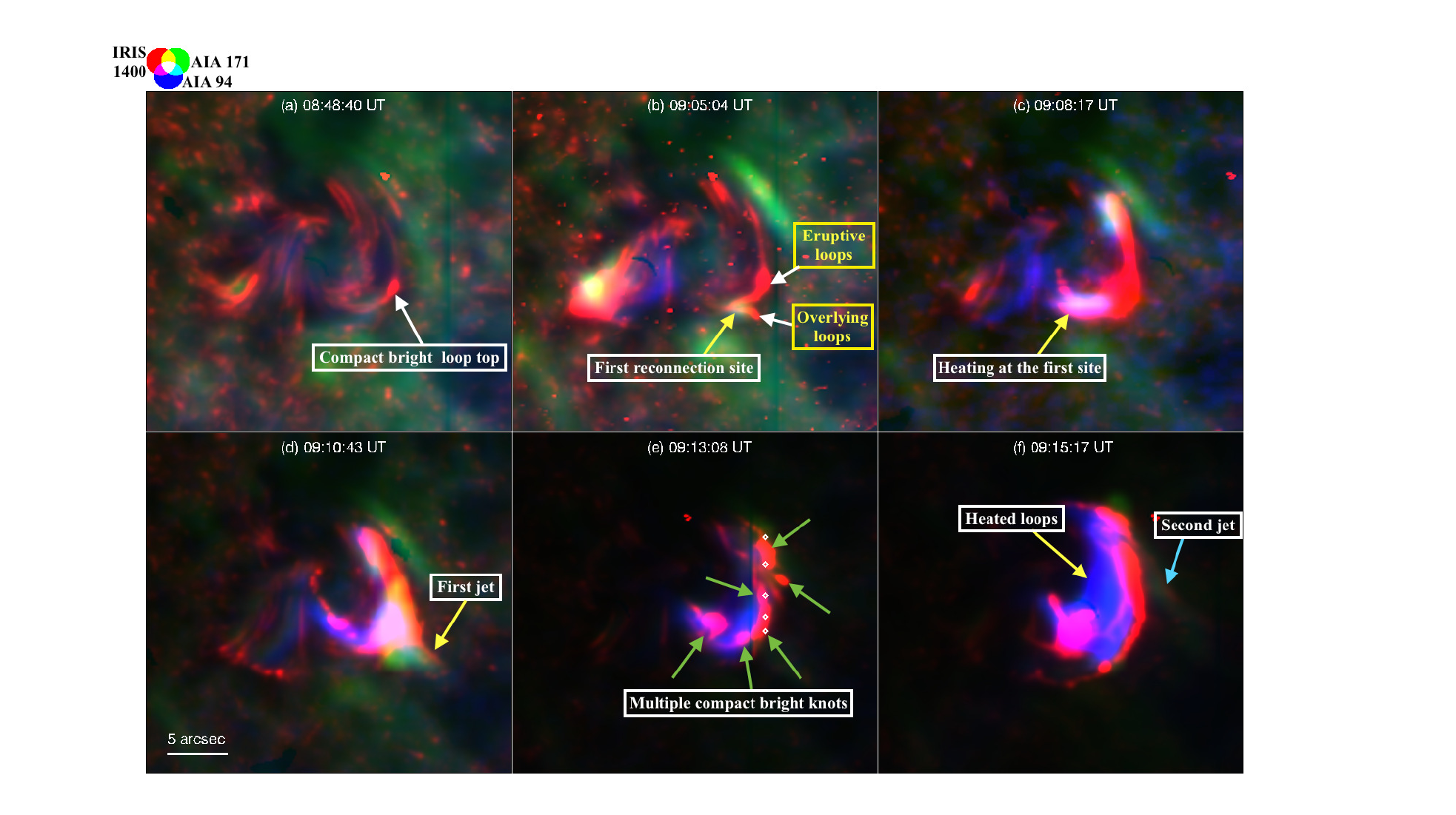}
\caption{The evolution of the spiral structure viewed in composite trio-temperature images. The red, green and blue components of the coloured images represent the IRIS 1400\,\AA\ ($8\times10^4$\,K), AIA 171\,\AA\ ($6.3\times10^5$\,K), and AIA 94\,\AA\ ($6.3\times10^6$\,K) channels, respectively. Panels (a) to (f) display six different snapshots of the event with the observing time denoted. Each image is shown in its own intensity scale in order to present the details of the structures. The diamond symbols in panel (e) denote where magnetic reconnection occurs as inferred by the explosive event spectra (see Figure\,\ref{figspimg}e). (An animation is provided).}
\label{fig3color}
\end{figure*}

The eruption was initiated by a small brightening occurring at a location where the cluster of bundled loops crosses another smaller loop system (Figure\,\ref{fig3color}b--c). This small activity is very likely a result of magnetic reconnection. This is evidenced by the heating in this region (Figure\,\ref{fig3color}c) as well as a small bright blob moving away from the crossing point toward the loop top (animation 2). The eruption of the event started while the moving blob reached the loop top (Figure\,\ref{fig3color} and animations 1--2).
This eruption generated upflows of a bulk of filament-like cool plasmas, which moved along the large loop at a speed of about 250\,km\;s$^{-1}$\ in the plane-of-sky (Figure\,\ref{fig3color}d and Figure\,\ref{figsv}). The ejected plasmas (hereafter, the first jet) were seen in IRIS and AIA channels corresponding to temperatures higher than 25\,000\,K, indicating a multi-thermal nature (see animation 1 around 09:11\,UT). 
We suggest that the reconnection released constraints of some over-lying loops, in agreement with the scenario proposed for coronal jets\,\citep{Pariat2009ApJ}.
The IRIS spectra showed clear signatures of rising cool plasmas co-spatial with the cluster of bundled loops. This can be seen from  absorption (dark) components embedded in the background spectra (see Figure\,\ref{figspimg}). The absorption components are shifted toward the short wavelength, indicating an upflow of cool plasmas at a velocity as large as 100\,km\;s$^{-1}$\ (Figure\,\ref{figspimg}). This is consistent with the imaging data.

\par
When the spiral structure was erupting, the O\,{\sc iv} 1399.8\,\AA\ and 1401.2\,\AA\ lines were strong enough (with a good signal-to-noise ratio) such that their intensity ratios could be used for electron density ($n_e$) diagnostics.  In Figure\,\ref{figsd}, we present the diagnostics results by using this method. From the model (Figure\,\ref{figsd}c) given by the CHIANTI atomic database\,\citep{Dere1997AAS,2013ApJ...763...86L}, the line ratio is sensitive to the electron density in a density range of $10^{9}$ to $10^{12.5}$ cm$^{-3}$ when line ratios are from 0.18 to 0.42. While many locations of the event have a line ratio exceeding 0.42 (Figure\,\ref{figsd}b), the electron density derived from this method gives an estimation at the lower limit (Figure\,\ref{figsd}d). While the event was eruptive, we find a lower limit of $\sim3\times10^{12}$\,cm$^{-3}$ for the eruptive loops (Figure\,\ref{figsd}d).

\begin{longtable}{p{0.2\linewidth} p{0.76\linewidth}}
\caption{Dynamic evolution of the active events.}
\label{tab_evolution}\\

09:03:59 UT&The first reconnection site appears as a small ($\lesssim$1$^{''}$) brightening at the cross-section between a small loop system and the main eruptive loops of the event. This brightening can be seen in all the IRIS and AIA passbands (except AIA UV) passbands, but can only be resolved in the AIA 94 \AA\ passband after 09:04:37 UT (i.e. ~40 s later than the others).\\

09:06:03 UT&A bright structure appears at the first reconnection site. It then splits into two blobs that are ejected in opposite directions including one moving toward the top of the main eruptive loops. The blob that moving toward the loop top reaches the original bright loop top around 09:06:35 UT. The moving motion of these two blobs can be clearly seen in IRIS SJ 1330 \AA\ and 1400 \AA\ data.\\

09:06:40 UT&The brightness of the loop top starts rapidly increasing, initially in the IRIS SJ 1330 \AA\ and 1400\,\AA\ images, with no clear response in the AIA EUV channels.\\

09:07:45 UT&While the brightness of the loop top keeps increasing in the IRIS 1330 \AA\ and 1400 \AA\ passbands, a bright feature appears at the first reconnection site in all AIA EUV passbands. This bright feature shows clear motion moving toward the loop top.\\

09:10:26 UT&The first plasma ejection that roots at the loop top is present. This ejection can be seen in all the channels including IRIS SJ and AIA UV and EUV passbands without any clear time lag (with the present cadence). \\

09:12:14 UT&The first plasma ejection and the brightness of the eruptive loops are slightly dimmed.\\

09:12:30 UT&The first plasma ejection continues evolving. Dark ejecting features appear at the north of its bright thread, and both dark and bright components of the ejections are rejected in the same direction. The dark component of the ejection is clearly seen in IRIS SJ 1400 \AA\ and AIA 304 \AA\ passbands, and is also captured by the IRIS spectral slit at about 09:13:18 UT. The brightness of the eruptive structure rapidly increases after this ejection.\\

09:14:28 UT&The eruption of the event reaches its peak intensity, and most AIA EUV channels are saturated. AIA 1700 \AA\ shows two bright features at the footpoints of the eruptive loops. However, there is no sign of plasma ejection at this stage of the eruption.\\

09:15:33 UT&The second plasma ejection is clearly seen. The initial thread is firstly ejected from a location about 5\,$^{''}$ north of the loop top (where the first plasma ejection was rooted). \\

09:16:05 UT&Many other threads of the second plasma ejection are clearly seen, especially in IRIS SJ 1400 \AA\ image. \\

09:17:58 UT& Down-flowing plasma can be seen in IRIS 1400 \AA. While also seen in the AIA 304 \AA, 171 \AA\ and 193 \AA\ passbands, the falling back features consist of both bright and dark components.\\

09:23:53 UT&Viewed in IRIS SJ images, the eruption process ends. The general structure of the feature is similar to its pre-eruption topology, suggesting the event was partially erupted and might only have changed the structure in localised regions. In the AIA 304 \AA, 171 \AA\ and 193 \AA\ passbands, we see that the down-falling plasma (with both dark and bright components) is continuing, lasting until 09:30:30 UT.\\

09:30:30 UT &The end of the eruption. The general structure of the feature is similar to the structure before the eruption.\\

\end{longtable}

\section{Signatures of magnetic braids}
By following the evolution of the event (Figure\,\ref{fig3color} and online animations 1--3), we found several pieces of observational evidence of magnetic braids in the cluster of bundled loops. Although other physical processes (e.g. kink instability) might also take place in this transition region loops, we believe magnetic braids are one major cause of the eruption.

\par
The first piece of evidence is that multiple bright knots of a few arcsecs in size spread along the loop during the eruption (Figure\,\ref{fig3color}e).
These bright knots occur at the locations where multiple loops are found to be crossing each other (see Figure\,\ref{figs1400}).
The localised nature of these bright knots indicates non-uniform heating along the loops. This suggests that the magnetic braiding could be present\,\citep{Parker1988ApJ, Schrijver2007ApJ, cirtain2013nature}.
The volume of each individual bright knots is estimated to be 1 arcsec cubed according to the high-resolution IRIS SJ images.
By applying the value of $n_e$ calculated in Sect. 2 to an electron-proton plasma we find a total mass of $\sim2\times10^{12}$\,g for one knot.
Even though the IRIS SJ images have a spatial resolution as high as 0.35$^{''}$,
the very compact brightening of the small-scale phenomena ($\sim$3 times of the resolution) does not allow the instrument to resolve the 
the fine structuring (including sub-pixel scale) therein due to the point spread function\,\citep{DPB2014IRIS}.
Therefore, the actual volume and mass of the knots maybe much smaller.

\par
\begin{figure*}
\includegraphics[clip,trim=0cm 0cm 0cm 0cm,width=\textwidth]{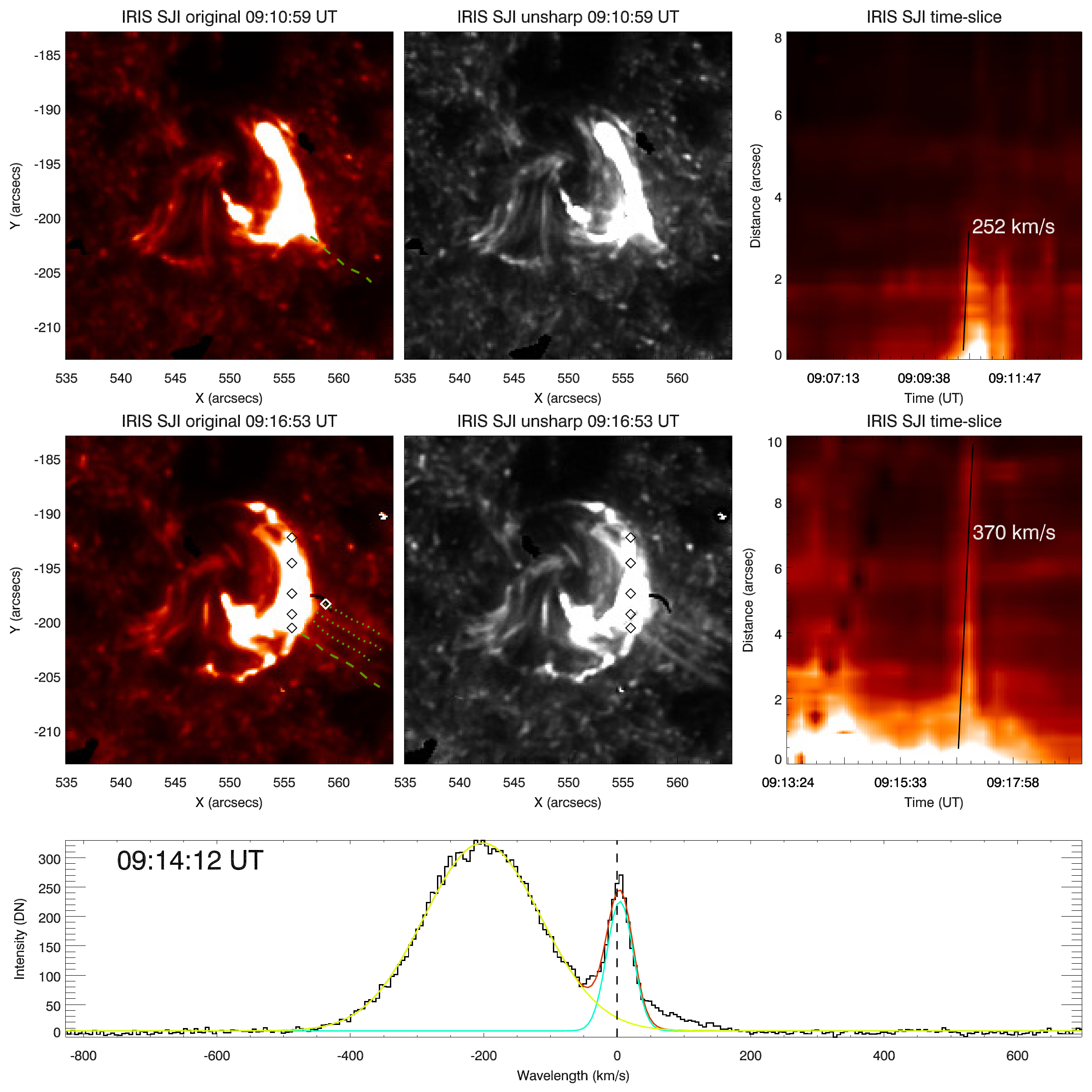}
\caption{ Properties of the jets observed during the eruption of the event.
Top row: the left panel shows the original IRIS SJ 1400\,\AA\ passband image at 09:10:59\,UT. The overlaid green dashed line marks the path of the first jet. The middle panel displays the same image processed using the unsharp mask technique to enhance the contrast of the image. The right-hand panel plots a space-time diagram calculated along the dashed green line from the left-hand panel. The solid black line indicates the linear fit used to calculate the velocity of the jet. 
The middle row plots the same as the top row but for the second series of jets (denoted by the green dashed and dotted lines in the left-hand panel).
The black diamonds in the central panel indicate the locations where magnetic reconnection at braiding sites occurred as determined in Figure\,3. These diamonds are in locations consistent with the foot-points of the jets. The space-time diagram in the right-hand panel is calculated for the dashed line in the left hand panel.
The bottom row plots a Si\,{\sc iv} 1403\,\AA\ spectrum taken at the base of the second jet (marked by a white diamond symbol in the left panel of the middle row) at the time shown in the panel, before that the jets are visible. A double Gaussian fit of the profile is given by the red line where the two Gaussian components are denoted by solid lines in yellow (contributed by the emission of the jet) and cyan (the background emission). The rest wavelength is denoted by the dashed line.}
\label{figsv}
\end{figure*}
\begin{figure*}
\includegraphics[clip,trim=0cm 0cm 0cm 0cm,width=\textwidth]{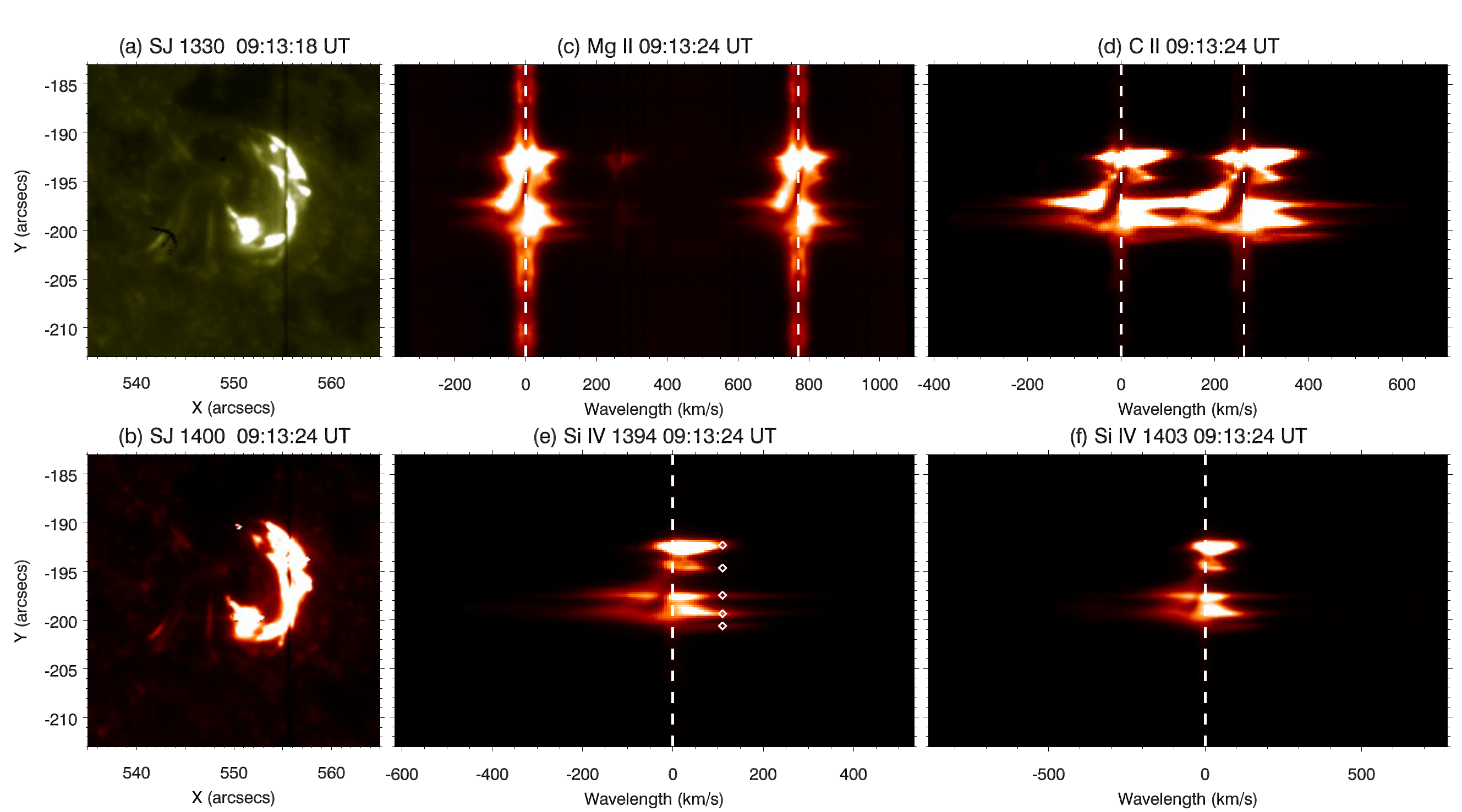}
\caption{The spectra of the spiral structure sampled during its eruption. (a) The event sampled by the IRIS SJ 1330\,\AA\ passband at the time six seconds prior to the presented spectra. (b) The event observed in IRIS SJ 1400\,\AA\ passband at the time when the spectra (c--f) were taken. The dark vertical line marks the location where the spectra were taken. The spectral images of the event along the spectral slit are given in (c): Mg\,{\sc ii}\,k\& h (representative of $1.4\times10^4$\,K), (d): C\,{\sc ii} 1334.6\,\AA\ and 1335.7\,\AA\ (representative of $4.2\times10^4$\,K), (e): Si\,{\sc iv}\,1393.9\,\AA\ and (f): Si\,{\sc iv}\,1402.8\,\AA\ (representative of $7.9\times10^4$\,K). The diamond symbols in panel (e) represent the locations (Y coordinates) of magnetic reconnection indicated by the explosive event spectra. The white dashed lines on the spectral images mark the rest wavelength of each line. (An animation is given online.)}
\label{figspimg}
\end{figure*}
\begin{figure*}
\includegraphics[clip,trim=0 0 8cm 0,width=\textwidth]{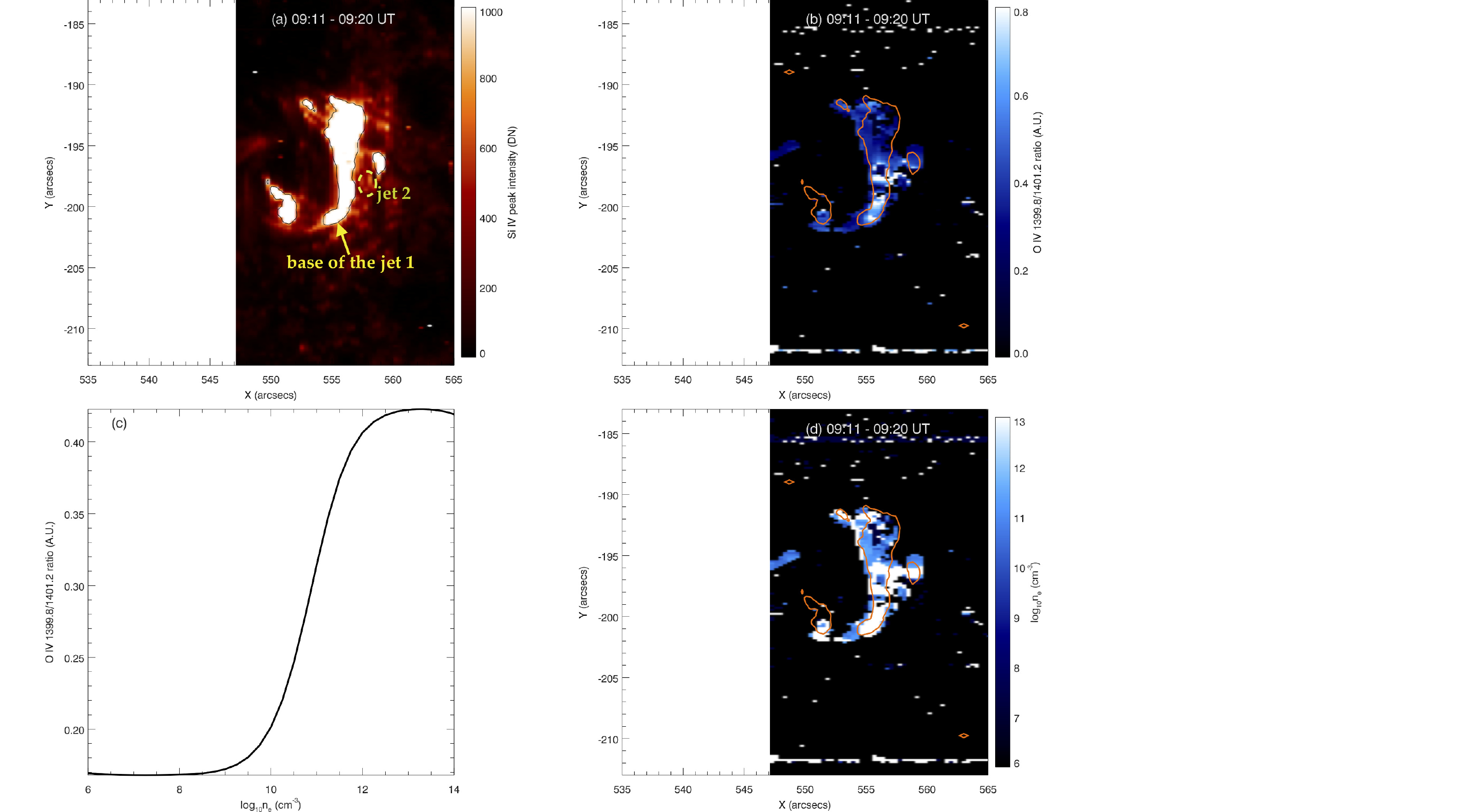}
\caption{The diagnostics of electron densities of the spiral structure. (a) The event sampled at the core of the Si IV 1403\,\AA\ line. The base of the first jet and the location of the second jet are denoted. The contour lines (in black) outline the eruptive structure. (b) The O\,{\sc iv} 1399.8\,\AA\ and 1401.2\,\AA\ line ratio map. The contour lines from the intensity map are overplotted in orange. (c) The relation between electron density and the O\,{\sc iv} 1399.8\,\AA\ and 1401.2\,\AA\ line ratio given by the CHIANTI database. (d) The derived electron density map of the field-of-view. Again the orange contours outline the eruptive structure.}
\label{figsd}
\end{figure*}

The second piece of evidence in support of the magnetic braiding scenario is the multi-threaded upward flows produced by the eruption.
About 5 minutes after the first jet, the second series of jets occurred (Figure\,\ref{fig3color}f and animation 1--2 and Figure\,\ref{figsv}).
These jets consisted of multiple threads of upflows that can be clearly resolved by the high-resolution IRIS slit-jaw (SJ) images (see Figure\,\ref{figsv} and animation 1).
These multi-threaded loops imply that the cluster of bundled loops is braided with the large loops at several distinct locations.
This is also supported by the fact that these jet threads are rooted at the same locations (presumably the braiding positions) where some of the bright knots are found .
The speed of one jet is measured as 370\,km\;s$^{-1}$\ in the plane of sky (Figure\,\ref{figsv}). 
Even though this value could have errors of $\sim$80\,km\;s$^{-1}$\ due to the low temporal resolution of the observations, 
the speed is consistent with the IRIS spectral observations that reveal an upward velocity of 200\,km\;s$^{-1}$\ along the line-of-sight (see animation 3 around 09:14:12\,UT and Figure\,\ref{figsv}). The high speed of this jet is in line with the magnetic reconnection scenario\,\citep{Tian2014Sci}, which would suggest outflows at the order of the Alfv\'en speed (inferred later in next section).
Since the plasma ejections are initiated in the transition region, if magnetic braids occur between closed and open field lines, they can supply some mass along open field that may further develop into the solar wind.

\par
The spectroscopic observations provided further evidence of magnetic braiding in the eruption. The transition region spectra of the bright knots displayed very broad profiles with strong intensity enhancements in both wings (animation 3). Such spectra are signatures of bi-directional flows\,\citep[e.g.][]{Brueckner1983ApJ,1989SoPh..123...41D,Innes1997nature}. They are so-called transition region explosive events. The most plausible scenario explaining the formation of explosive events is magnetic reconnection\,\citep[e.g.][]{Brueckner1983ApJ, Dere1991JGR, Innes1997nature, Peter2014Sci, Huang2014ApJ, 2015ApJ...810...46H,2016ApJ...824...96T,Huang2017mnras}.
The localisation of the explosive events indicates a localised topology of reconnection.
This is consistent with the overall picture of magnetic braiding\,\citep{Parker1988ApJ}. In Figure\,\ref{figspimg}, we present several images of the spectral slit taken for the event.  The explosive events are present only in locations of magnetic braids as determined from the imaging data\,(see Figure\,\ref{fig3color}e). They are also in agreement with the footpoints of the jets (see Figure\,\ref{figsv}). 
These observations indicate that the magnetic reconnection occurred at places where the field lines were braided.

\section{Three-dimensional field extrapolation}
\begin{figure*}
\includegraphics[width=\textwidth]{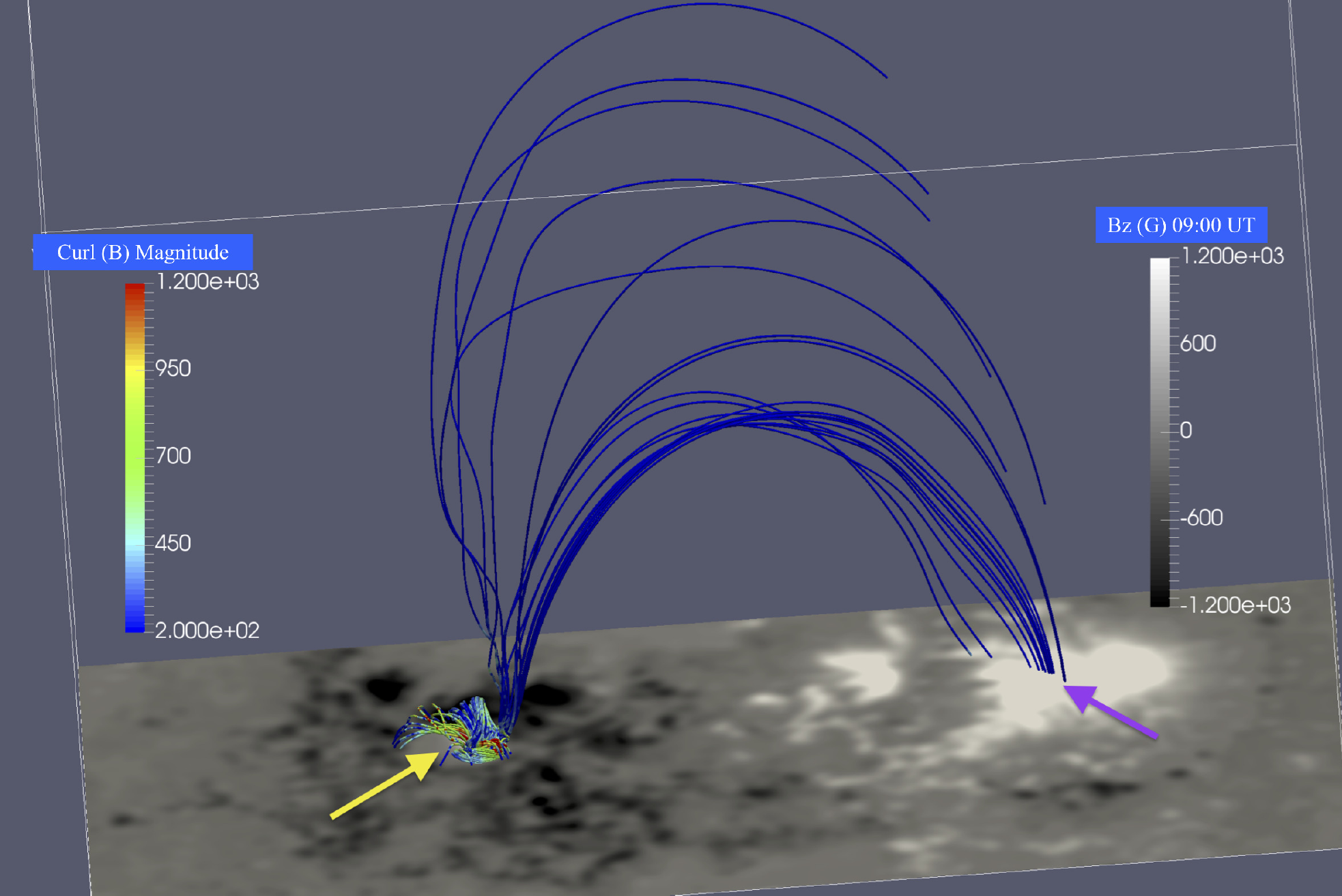}
\caption{The 3D magnetic field of the full field-of-view used in the extrapolation. The underlying image in grey-scale displays the magnetogram of the region. The color of the field lines are representative of the curl values of the magnetic field. The location of the eruption studied in this paper is denoted by the arrow coloured in yellow. A class of larger field lines representative of the large loops shown in Fig.\,\ref{fig193_large} connect the event and a remote region marked by the purple arrow.}
\label{figsk}
\end{figure*}

\begin{figure*}
\includegraphics[clip,trim=0cm 0cm 0cm 0cm,width=0.95\textwidth]{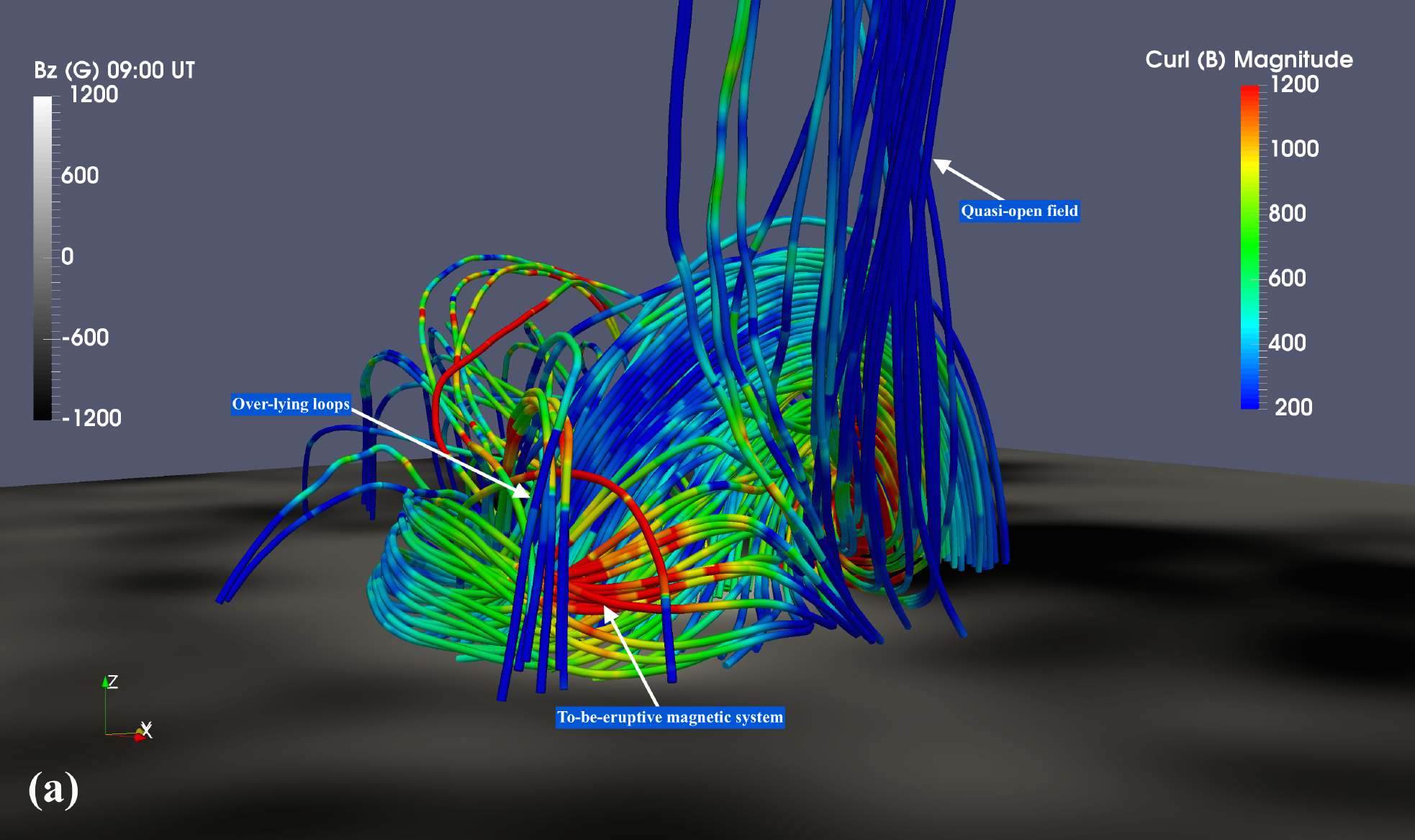}\\
\includegraphics[clip,trim=0cm 0cm 0cm 0cm,width=0.95\textwidth]{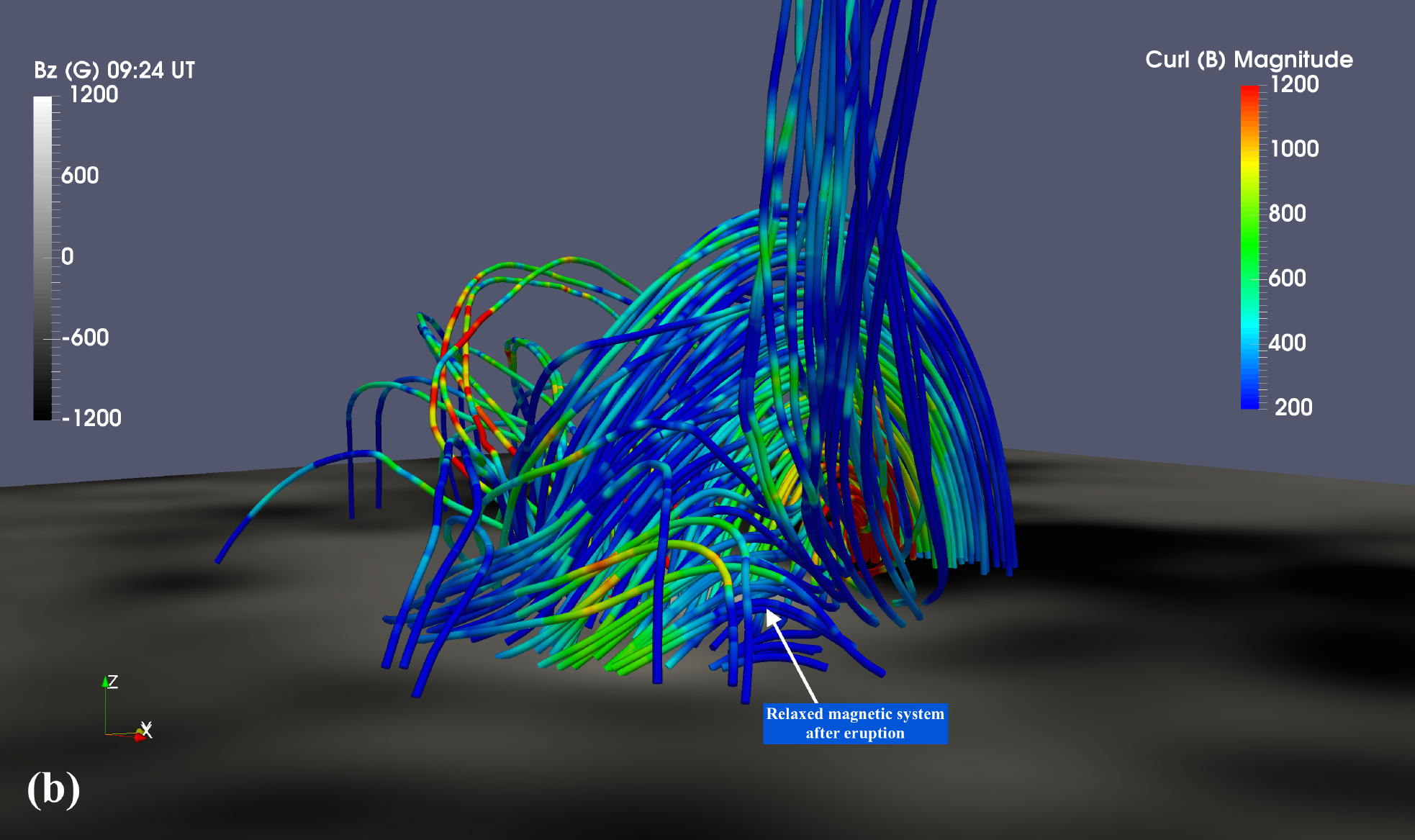}
\caption{The extrapolated magnetic field of the spiral structure before its eruption at 09:00\,UT (panel a) and after its eruption at 09:24\,UT (panel b). The underlying grey-scaled image displays the line-of-sight component of the magnetic field of the region (Bz). The field line colour is coded with the magnitude of the curl of the magnetic field, which is proportional to the electric current density in the region around the field lines. (An animation is provided.)}
\label{fig3dfield}
\end{figure*}

In order to obtain the three-dimensional (3D) magnetic topology of the event, we applied a non-linear force-free field (NLFFF) extrapolation model\,\citep{wiegelmann2007sol} on vector magnetic data obtained with HMI and pre-calibrated by the instrument team. The vector magnetic field data use the cylindrical equal area projection (CEA) coordinates\,\citep{sun2013arxiv,Sun2017ApJ}. The field-of-view for which the extrapolation was computed includes the whole Active Region (shown in Figure\,\ref{figsk}). Figure\,\ref{fig3dfield} displays a cut-out of the entire extrapolated region, which was selected to highlight the detail of the magnetic field of the eruptive event.

\par
The 3D magnetic field of the region shows that the spiral arms of the spiral structure comprise multiple strands. One of the arms corresponds to the eruptive bundle of loops  that was trapped by an over-lying loop system. 
The 3D magnetic field reveals another loop system that is of a much larger scale (see Figures\,\ref{figsk}\&\ref{fig3dfield}).
This larger loop system is rooted in the eruptive region and connects to a remote footpoint. It corresponds to the loops that channel the observed jets, and is also seen in AIA 193\,\AA\ images in the form of typical coronal loops (Fig.\,\ref{fig193_large}).
Since the loop system is much larger than the eruptive loops, we consider it as quasi-open.

\par
The curls of the vector magnetic field ($\nabla\times\overrightharp{B}$) along the eruptive loops vary significantly (Figure\,\ref{fig3dfield}).
This implies a complex winding nature of this loop system, because the curl is representative of rotation of the field lines in a local region (determined by the sampling of the data used for the calculus of difference).
The curl of the magnetic field is much larger at the loop top, corresponding to a compact bright knot seen in the IRIS SJ observations (see the online animation). 
Because the curl is proportional to the electric current, a possible interpretation is that the bright knot is representative of a place of strong current, providing an appropriate condition of magnetic reconnection\,\citep{priest2000book}.

\par
Using the extrapolated data, the Alfv\'en speed ($v_A$) at the footpoint of the jets can be obtained by $v_A={B}/{\sqrt{4\pi\rho}}$ in cgs units, where $B$ is the magnetic strength and $\rho$ is the density. In our case, $B$ is given as 300\,G based on the magnetic field extrapolation and $\rho$ is set to be $5\times10^{-12}$\,g\,cm$^{-3}$ by applying the above value of $n_e$ to an electron-proton plasma.
The accuracy of the calculation relies on the measurements of the magnetic strength and the density, which could bring in some uncertainty in our case. Based on our measurements, we obtain an estimate of 380\,km\;s$^{-1}$\ for the Alfv\'en speed at the footpoint of the jets, comparable to the speed of the jets observed presently. This is consistent with the general picture of magnetic reconnection that suggests an outflow at the oder of the Alfv\'en speed\,\citep{priest2000book}.

\par
After the reconnection, the loop system is relaxed and the curl values have been reduced (Figure\,\ref{fig3dfield} and animation 4). 
This suggests that the eruption of the twisted flux rope expels the twist from beneath\,\citep{2016ApJ...827....4W,2017A&A...604A..76M}.
Current dissipation during this process could heat the local plasmas due to the resistivity therein\,\citep{priest2000book}. 
As seen in the AIA 94\,\AA\ channel, the plasmas have been heated to at least 6.3\,MK (the feature coloured by blue, better viewed in Figure\,\ref{fig3color}f).
This agrees with the simulations presented in previous studies\,\citep{WilmotSmith2010AA}.

\par
Based on the extrapolated 3D field, we calculate the component of the electric current parallel to the magnetic field ($j_\parallel$) by ignoring the displacement current, i.e. 
$$j_\parallel=\frac{c}{4\pi}\frac{\overrightharp{B}}{|\overrightharp{B}|}\cdot(\nabla\times\overrightharp{B}),$$ where $c$ is the speed of light, $\overrightharp{B}$ is the vector of the magnetic strength. The squashing factor ($Q$) and twist number were further derived from the NLFFF results\,\citep{Liu2016ApJ}. In Figure\,\ref{figqtw}, we display the IRIS SJ 1400\,\AA\ image, and the maps of the $j_\parallel$,  $Q$ and twist number of the region in the CEA coordinates. The high $Q$ values shown on the $Q$ map indicate the locations of three-dimensional quasi-separatrix layers (QSLs) at the given height. The twist number measures how many turns two infinitesimally close field lines wind about each other\,\citep{Liu2016ApJ}. 
The presence of opposite direction of twists at the same location is a hint of multiple magnetic braids among more than two loop threads.
Strong parallel currents are found in this eruption (Figure\,\ref{figqtw}), which are consistent with the simulations of magnetic braids\,\citep{WilmotSmith2009ApJ,WilmotSmith2010AA}.
The event is also positioned at regions of large squashing factors and twist number (see Figure\,\ref{figqtw}). This indicates the presence of quasi-separatrix layers (QSLs) and possible braiding of {the} magnetic field\,\citep{Liu2016ApJ}. 
The explosive events identified in the IRIS spectral data are located in regions where the parallel current, $Q$ value and twist number are relatively high. 
These analyses are in line with the scenario of magnetic reconnection in magnetic braids\,\citep{WilmotSmith2009ApJ}.

\par
The null points were identified by a procedure\,\citep{Liu2016NatSR} based on Newton-Raphson method. 
Five nulls are identified, which are found to group together at a location in the vicinity of the eruptive region  and about 2$^{''}$ above the photosphere. 
Their projections on the photosphere are shown in Figure\,\ref{figqtw}. 
Considering the overall topology of the Q map showing a circular structure with internal ridge of high Q,
this is a classic characteristic associated with the spine and fan structures of a coronal null point\,\citep{2009ApJ...700..559M,2016SoPh..291.1739P}.
The clustering of null points here could be a result of some instability\,\citep{2014PhPl...21j2102W}. The first jet observed in the event here could be the result of null point reconnection.

\par
The geometry distribution of the photospheric magnetic flux of this event is similar to some other eruptive phenomena in the solar atmosphere, and most are associated with spine field lines and (quasi-)separatrix dome\,\citep[e.g.][]{2013ApJ...778..139S,2017A&A...604A..76M}. 
In such the magnetic field topologies, plasmoids produced by the magnetic reconnection can introduce a very complex field topology including twisting and braiding,
and the localized bright knots in the eruptive phenomena might be associated with unsteady ``breakout'' reconnection\,\citep{2013ApJ...778..139S} of a flux rope from beneath the null point's separatrix dome\,\citep{2016ApJ...827....4W}.
Similar processes cannot be ruled out in the event studied here, but the observations do not allow one to judge whether the magnetic braids were the causes or the results of the reconnection.

\section{Conclusion and Discussion}
In the present study, we report on multi-instrument observations of eruptions that occur in an arm of a spiral structure in the solar atmosphere.
The eruptive part of the structure consists of multiple loop threads resolved in the high-resolution IRIS SJ images.
The loop threads appear to join in the same location where the most intense eruption occurred, providing a favorable environment for braiding.
The eruption produced two series of jets that origin in the transition region, and thus potentially supplies mass to the corona.
One ejected cool filament plasma with an apparent speed of 250\,km\;s$^{-1}$, and the other one consists of well-separated multiple threads with an apparent speed about 370\,km\;s$^{-1}$, which is comparable to the Alfv\'en speeds estimated in the footpoint of the jets.

\par
During the eruption, several pieces of evidence of magnetic braiding are revealed, including localized bright knots, multiple jet threads and transition region explosive events.
The bright knots, footpoints of the jet threads and transition region explosive events fall into the same locations within the eruptive structures, and
this strongly indicates localised magnetic reconnection events in line with the magnetic braiding scenario.
This scenario is also supported by the 3D magnetic topology of the event.
The 3D extrapolated field reveals a complex twisted magnetic system, in which
the curl of the field along the eruptive loops vary significantly. This indicates a complex winding nature of the system.
The eruptive event was positioned at a region with 
strong parallel currents, large squashing factors and twist number, which indicates the presence of quasi-separatrix layers and possible braiding of magnetic field.
While these are convincing evidence in favor of magnetic braiding in the eruption, it requires higher-temporal-resolution observation to determine whether the magnetic braids were the causes or the results of the reconnection, and higher-spatial-resolution data to resolve the details in the braiding geometry.

\par
While magnetic-braid-related reconnection is one of the important candidate mechanisms for heating the solar corona\,\citep{Parker1983ApJb, Parker1988ApJ, Klimchuk2006SoPh, Schrijver2007ApJ, cirtain2013nature}, the observations presented here demonstrate that they might also play a key role in localised energy release during solar eruptions. The magnetic braids, mostly driven by the random motions on the photosphere\,\citep{Parker1983ApJb}, 
could build up magnetic energy stored in magnetic structures of the solar atmosphere (e.g. loops, filaments). This could then be released during an eruption.
While it is released via magnetic reconnection, 
it can provide energy to heat and accelerate plasmas\,\citep{Parker1988ApJ}.
Since the photosphere is constantly buffeted by underlying convections\,\citep{Parker1983ApJb, Parker1988ApJ}, magnetic braids should be prevalent throughout the solar atmosphere. 
Although most magnetic braids may not be resolved by present instruments, their relevant processes 
could be part of the mechanisms to understand the three major solar physics problems (i.e., the formation of solar eruptions, coronal heating, and the origin of the solar wind).

\par
{\it Acknowledgement:}
We greatly acknowledge many constructive comments given by the anonymous referees. This research is supported by the National Natural Science Foundation of China (41474150, 41627806, 41574166, 41404135). IRIS is a NASA small explorer mission developed and operated by LMSAL with mission operations executed at NASA Ames Research center and major contributions to downlink communications funded by ESA and the Norwegian Space Center. Courtesy of NASA/SDO, the AIA and HMI science teams and JSOC. ParaView has been used to visualised the magnetic field. Z.H. acknowledges M. Madjarska for encouragement and the Young Scholar program of Shandong University, Weihai (2017WHWLJH07). C.J.N. is thankful to the STFC for the support received to conduct this research. H.T. thanks the Max Planck Partner Group program. T.W. acknowledges DFG-grant WI 3211/4-1.

{\color{red}
\begin{figure*}
\includegraphics[clip,trim=0cm 2.6cm 0cm 1cm,width=\textwidth]{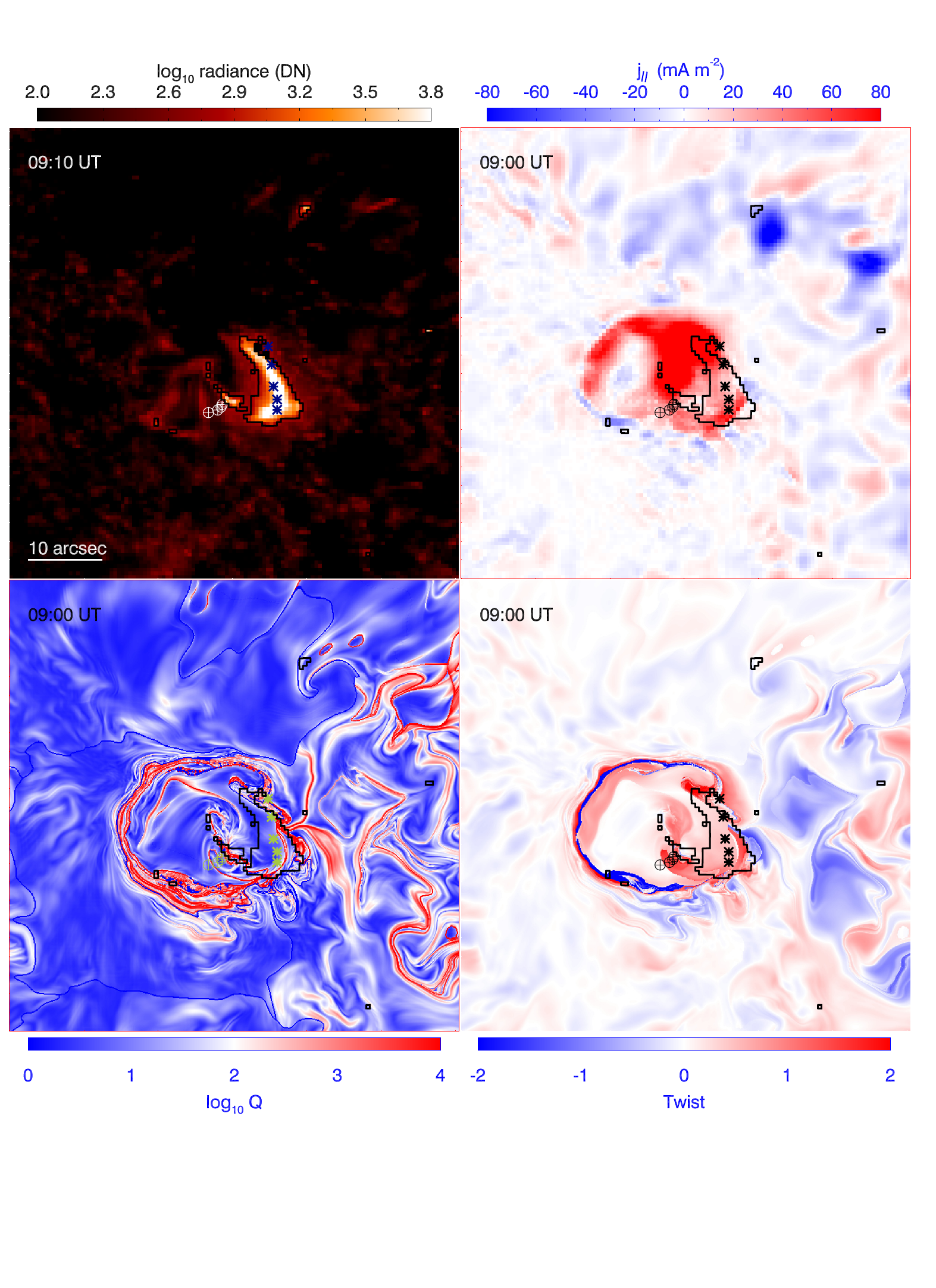}
\caption{The eruptive event in IRIS SJ 1400\,\AA\ passband at 09:10\,UT (top-left) and the maps of parallel current (top-right), the squashing factor $Q$ (bottom-left) and twist number (bottom-right) in the photosphere deduced from the extrapolated fields at 09:00\,UT. The SJ 1400\,\AA\ image has been downgraded to the HMI resolution and translated to the cylindrical equal area projection (CEA) coordinates that are used in the HMI vector magnetic field data and therefore in the images presented in the other panels here. The contours are representative of the emission levels of the event seen in IRIS SJ 1400\,\AA. The asterisks mark the locations where the explosive events occurred. The circles with crosses indicate the locations above which the null points are found.}
\label{figqtw}
\end{figure*}
}

\bibliographystyle{aasjournal}
\bibliography{bibliography}

\begin{thebibliography}{}
\expandafter\ifx\csname natexlab\endcsname\relax\def\natexlab#1{#1}\fi

\bibitem[{{Antiochos} {et~al.}(2011){Antiochos}, {Miki{\'c}}, {Titov},
  {Lionello}, \& {Linker}}]{Antiochos2011ApJ}
{Antiochos}, S.~K., {Miki{\'c}}, Z., {Titov}, V.~S., {Lionello}, R., \&
  {Linker}, J.~A. 2011, \apj, 731, 112

\bibitem[{{Brueckner} \& {Bartoe}(1983)}]{Brueckner1983ApJ}
{Brueckner}, G.~E., \& {Bartoe}, J.-D.~F. 1983, \apj, 272, 329

\bibitem[{Chen(2011)}]{Chen2011lrsp}
Chen, P.~F. 2011, Living Reviews in Solar Physics, 8, doi:10.1007/lrsp-2011-1

\bibitem[{{Cirtain} {et~al.}(2013){Cirtain}, {Golub}, {Winebarger}, {de
  Pontieu}, {Kobayashi}, {Moore}, {Walsh}, {Korreck}, {Weber}, {McCauley},
  {Title}, {Kuzin}, \& {Deforest}}]{cirtain2013nature}
{Cirtain}, J.~W., {Golub}, L., {Winebarger}, A.~R., {et~al.} 2013, \nat, 493,
  501

\bibitem[{{Crooker} {et~al.}(2012){Crooker}, {Antiochos}, {Zhao}, \&
  {Neugebauer}}]{Crooker2012JGRA}
{Crooker}, N.~U., {Antiochos}, S.~K., {Zhao}, X., \& {Neugebauer}, M. 2012,
  Journal of Geophysical Research (Space Physics), 117, A04104

\bibitem[{{De Pontieu} {et~al.}(2014{\natexlab{a}}){De Pontieu}, {Rouppe van
  der Voort}, {McIntosh}, {Pereira}, {Carlsson}, {Hansteen}, {Skogsrud},
  {Lemen}, {Title}, {Boerner}, {Hurlburt}, {Tarbell}, {Wuelser}, {De Luca},
  {Golub}, {McKillop}, {Reeves}, {Saar}, {Testa}, {Tian}, {Kankelborg},
  {Jaeggli}, {Kleint}, \& {Martinez-Sykora}}]{DP2014Sci}
{De Pontieu}, B., {Rouppe van der Voort}, L., {McIntosh}, S.~W., {et~al.}
  2014{\natexlab{a}}, Science, 346, 1255732

\bibitem[{{De Pontieu} {et~al.}(2014{\natexlab{b}}){De Pontieu}, {Title},
  {Lemen}, {Kushner}, {Akin}, {Allard}, {Berger}, {Boerner}, {Cheung}, {Chou},
  {Drake}, {Duncan}, {Freeland}, {Heyman}, {Hoffman}, {Hurlburt}, {Lindgren},
  {Mathur}, {Rehse}, {Sabolish}, {Seguin}, {Schrijver}, {Tarbell},
  {W{\"u}lser}, {Wolfson}, {Yanari}, {Mudge}, {Nguyen-Phuc}, {Timmons}, {van
  Bezooijen}, {Weingrod}, {Brookner}, {Butcher}, {Dougherty}, {Eder},
  {Knagenhjelm}, {Larsen}, {Mansir}, {Phan}, {Boyle}, {Cheimets}, {DeLuca},
  {Golub}, {Gates}, {Hertz}, {McKillop}, {Park}, {Perry}, {Podgorski},
  {Reeves}, {Saar}, {Testa}, {Tian}, {Weber}, {Dunn}, {Eccles}, {Jaeggli},
  {Kankelborg}, {Mashburn}, {Pust}, {Springer}, {Carvalho}, {Kleint}, {Marmie},
  {Mazmanian}, {Pereira}, {Sawyer}, {Strong}, {Worden}, {Carlsson}, {Hansteen},
  {Leenaarts}, {Wiesmann}, {Aloise}, {Chu}, {Bush}, {Scherrer}, {Brekke},
  {Martinez-Sykora}, {Lites}, {McIntosh}, {Uitenbroek}, {Okamoto}, {Gummin},
  {Auker}, {Jerram}, {Pool}, \& {Waltham}}]{DPB2014IRIS}
{De Pontieu}, B., {Title}, A.~M., {Lemen}, J.~R., {et~al.} 2014{\natexlab{b}},
  \solphys, 289, 2733

\bibitem[{{Dere} {et~al.}(1989){Dere}, {Bartoe}, \&
  {Brueckner}}]{1989SoPh..123...41D}
{Dere}, K.~P., {Bartoe}, J.-D.~F., \& {Brueckner}, G.~E. 1989, \solphys, 123,
  41

\bibitem[{{Dere} {et~al.}(1991){Dere}, {Bartoe}, {Brueckner}, {Ewing}, \&
  {Lund}}]{Dere1991JGR}
{Dere}, K.~P., {Bartoe}, J.-D.~F., {Brueckner}, G.~E., {Ewing}, J., \& {Lund},
  P. 1991, \jgr, 96, 9399

\bibitem[{{Dere} {et~al.}(1997){Dere}, {Landi}, {Mason}, {Monsignori Fossi}, \&
  {Young}}]{Dere1997AAS}
{Dere}, K.~P., {Landi}, E., {Mason}, H.~E., {Monsignori Fossi}, B.~C., \&
  {Young}, P.~R. 1997, \aaps, 125, doi:10.1051/aas:1997368

\bibitem[{{Fisk}(2003)}]{Fisk2003JGR}
{Fisk}, L.~A. 2003, Journal of Geophysical Research (Space Physics), 108, 1157

\bibitem[{{Hong} {et~al.}(2017){Hong}, {Jiang}, {Yang}, {Li}, \&
  {Xu}}]{2017ApJ...835...35H}
{Hong}, J., {Jiang}, Y., {Yang}, J., {Li}, H., \& {Xu}, Z. 2017, \apj, 835, 35

\bibitem[{{Hong} {et~al.}(2016){Hong}, {Jiang}, {Yang}, {Yang}, {Xu}, \&
  {Xiang}}]{2016ApJ...830...60H}
{Hong}, J., {Jiang}, Y., {Yang}, J., {et~al.} 2016, \apj, 830, 60

\bibitem[{{Huang} {et~al.}(2017){Huang}, {Madjarska}, {Scullion}, {Xia},
  {Doyle}, \& {Ray}}]{Huang2017mnras}
{Huang}, Z., {Madjarska}, M.~S., {Scullion}, E.~M., {et~al.} 2017, \mnras, 464,
  1753

\bibitem[{{Huang} {et~al.}(2014){Huang}, {Madjarska}, {Xia}, {Doyle},
  {Galsgaard}, \& {Fu}}]{Huang2014ApJ}
{Huang}, Z., {Madjarska}, M.~S., {Xia}, L., {et~al.} 2014, \apj, 797, 88

\bibitem[{{Huang} {et~al.}(2015){Huang}, {Xia}, {Li}, \&
  {Madjarska}}]{2015ApJ...810...46H}
{Huang}, Z., {Xia}, L., {Li}, B., \& {Madjarska}, M.~S. 2015, \apj, 810, 46

\bibitem[{{Innes} {et~al.}(1997){Innes}, {Inhester}, {Axford}, \&
  {Wilhelm}}]{Innes1997nature}
{Innes}, D.~E., {Inhester}, B., {Axford}, W.~I., \& {Wilhelm}, K. 1997, \nat,
  386, 811

\bibitem[{{Kamio} {et~al.}(2010){Kamio}, {Curdt}, {Teriaca}, {Inhester}, \&
  {Solanki}}]{2010A&A...510L...1K}
{Kamio}, S., {Curdt}, W., {Teriaca}, L., {Inhester}, B., \& {Solanki}, S.~K.
  2010, \aap, 510, L1

\bibitem[{{Klimchuk}(2006)}]{Klimchuk2006SoPh}
{Klimchuk}, J.~A. 2006, \solphys, 234, 41

\bibitem[{{Landi} {et~al.}(2013){Landi}, {Young}, {Dere}, {Del Zanna}, \&
  {Mason}}]{2013ApJ...763...86L}
{Landi}, E., {Young}, P.~R., {Dere}, K.~P., {Del Zanna}, G., \& {Mason}, H.~E.
  2013, \apj, 763, 86

\bibitem[{{Lemen} {et~al.}(2012){Lemen}, {Title}, {Akin}, {Boerner}, {Chou},
  {Drake}, {Duncan}, {Edwards}, {Friedlaender}, {Heyman}, {Hurlburt}, {Katz},
  {Kushner}, {Levay}, {Lindgren}, {Mathur}, {McFeaters}, {Mitchell}, {Rehse},
  {Schrijver}, {Springer}, {Stern}, {Tarbell}, {Wuelser}, {Wolfson}, {Yanari},
  {Bookbinder}, {Cheimets}, {Caldwell}, {Deluca}, {Gates}, {Golub}, {Park},
  {Podgorski}, {Bush}, {Scherrer}, {Gummin}, {Smith}, {Auker}, {Jerram},
  {Pool}, {Soufli}, {Windt}, {Beardsley}, {Clapp}, {Lang}, \&
  {Waltham}}]{Lemen2012aia}
{Lemen}, J.~R., {Title}, A.~M., {Akin}, D.~J., {et~al.} 2012, \solphys, 275, 17

\bibitem[{{Li} {et~al.}(2016){Li}, {Zhang}, {Peter}, {Priest}, {Chen}, {Guo},
  {Chen}, \& {Mackay}}]{Li2016NatPh}
{Li}, L., {Zhang}, J., {Peter}, H., {et~al.} 2016, Nature Physics, 12, 847

\bibitem[{{Li} {et~al.}(2012){Li}, {Morgan}, {Leonard}, \&
  {Jeska}}]{2012ApJ...752L..22L}
{Li}, X., {Morgan}, H., {Leonard}, D., \& {Jeska}, L. 2012, \apjl, 752, L22

\bibitem[{{Liu} {et~al.}(2016{\natexlab{a}}){Liu}, {Chen}, {Wang}, \&
  {Liu}}]{Liu2016NatSR}
{Liu}, R., {Chen}, J., {Wang}, Y., \& {Liu}, K. 2016{\natexlab{a}}, Scientific
  Reports, 6, 34021

\bibitem[{{Liu} {et~al.}(2016{\natexlab{b}}){Liu}, {Kliem}, {Titov}, {Chen},
  {Wang}, {Wang}, {Liu}, {Xu}, \& {Wiegelmann}}]{Liu2016ApJ}
{Liu}, R., {Kliem}, B., {Titov}, V.~S., {et~al.} 2016{\natexlab{b}}, \apj, 818,
  148

\bibitem[{{Masson} {et~al.}(2009){Masson}, {Pariat}, {Aulanier}, \&
  {Schrijver}}]{2009ApJ...700..559M}
{Masson}, S., {Pariat}, E., {Aulanier}, G., \& {Schrijver}, C.~J. 2009, \apj,
  700, 559

\bibitem[{{Masson} {et~al.}(2017){Masson}, {Pariat}, {Valori}, {Deng}, {Liu},
  {Wang}, \& {Reid}}]{2017A&A...604A..76M}
{Masson}, S., {Pariat}, {\'E}., {Valori}, G., {et~al.} 2017, \aap, 604, A76

\bibitem[{{Pariat} {et~al.}(2009){Pariat}, {Antiochos}, \&
  {DeVore}}]{Pariat2009ApJ}
{Pariat}, E., {Antiochos}, S.~K., \& {DeVore}, C.~R. 2009, \apj, 691, 61

\bibitem[{{Parker}(1983{\natexlab{a}})}]{Parker1983ApJb}
{Parker}, E.~N. 1983{\natexlab{a}}, \apj, 264, 642

\bibitem[{{Parker}(1983{\natexlab{b}})}]{Parker1983ApJa}
---. 1983{\natexlab{b}}, \apj, 264, 635

\bibitem[{{Parker}(1988)}]{Parker1988ApJ}
---. 1988, \apj, 330, 474

\bibitem[{{Pesnell} {et~al.}(2012){Pesnell}, {Thompson}, \&
  {Chamberlin}}]{Pesnell2012sdo}
{Pesnell}, W.~D., {Thompson}, B.~J., \& {Chamberlin}, P.~C. 2012, \solphys,
  275, 3

\bibitem[{{Peter} {et~al.}(2014){Peter}, {Tian}, {Curdt}, {Schmit}, {Innes},
  {De Pontieu}, {Lemen}, {Title}, {Boerner}, {Hurlburt}, {Tarbell}, {Wuelser},
  {Mart{\'{\i}}nez-Sykora}, {Kleint}, {Golub}, {McKillop}, {Reeves}, {Saar},
  {Testa}, {Kankelborg}, {Jaeggli}, {Carlsson}, \& {Hansteen}}]{Peter2014Sci}
{Peter}, H., {Tian}, H., {Curdt}, W., {et~al.} 2014, Science, 346, 1255726

\bibitem[{{Pontin} {et~al.}(2016){Pontin}, {Galsgaard}, \&
  {D{\'e}moulin}}]{2016SoPh..291.1739P}
{Pontin}, D., {Galsgaard}, K., \& {D{\'e}moulin}, P. 2016, \solphys, 291, 1739

\bibitem[{{Pontin} {et~al.}(2017){Pontin}, {Janvier}, {Tiwari}, {Galsgaard},
  {Winebarger}, \& {Cirtain}}]{Pontin2017ApJ}
{Pontin}, D.~I., {Janvier}, M., {Tiwari}, S.~K., {et~al.} 2017, \apj, 837, 108

\bibitem[{{Priest} \& {Forbes}(2000)}]{priest2000book}
{Priest}, E., \& {Forbes}, T. 2000, {Magnetic Reconnection: MHD theory and
  applications} (Cambridge, UK: Cambridge University Press), 612

\bibitem[{{Priest} {et~al.}(1998){Priest}, {Foley}, {Heyvaerts}, {Arber},
  {Culhane}, \& {Acton}}]{Priest1998Nature}
{Priest}, E.~R., {Foley}, C.~R., {Heyvaerts}, J., {et~al.} 1998, \nat, 393, 545

\bibitem[{{Schou} {et~al.}(2012){Schou}, {Scherrer}, {Bush}, {Wachter},
  {Couvidat}, {Rabello-Soares}, {Bogart}, {Hoeksema}, {Liu}, {Duvall}, {Akin},
  {Allard}, {Miles}, {Rairden}, {Shine}, {Tarbell}, {Title}, {Wolfson},
  {Elmore}, {Norton}, \& {Tomczyk}}]{Schou2012hmi}
{Schou}, J., {Scherrer}, P.~H., {Bush}, R.~I., {et~al.} 2012, \solphys, 275,
  229

\bibitem[{{Schrijver}(2007)}]{Schrijver2007ApJ}
{Schrijver}, C.~J. 2007, \apjl, 662, L119

\bibitem[{{Schrijver} {et~al.}(1998){Schrijver}, {Title}, {Harvey}, {Sheeley},
  {Wang}, {van den Oord}, {Shine}, {Tarbell}, \&
  {Hurlburt}}]{Schrijver1998Nature}
{Schrijver}, C.~J., {Title}, A.~M., {Harvey}, K.~L., {et~al.} 1998, \nat, 394,
  152

\bibitem[{Shibata \& Magara(2011)}]{Shibata2011lrsp}
Shibata, K., \& Magara, T. 2011, Living Reviews in Solar Physics, 8,
  doi:10.1007/lrsp-2011-6

\bibitem[{{Sterling} {et~al.}(2015){Sterling}, {Moore}, {Falconer}, \&
  {Adams}}]{Sterling2015Nature}
{Sterling}, A.~C., {Moore}, R.~L., {Falconer}, D.~A., \& {Adams}, M. 2015,
  \nat, 523, 437

\bibitem[{Su {et~al.}(2013)Su, Veronig, Holman, Dennis, Wang, Temmer, \&
  Gan}]{Su2013}
Su, Y., Veronig, A.~M., Holman, G.~D., {et~al.} 2013, Nature Physics, 9, 489

\bibitem[{{Su} {et~al.}(2012){Su}, {Wang}, {Veronig}, {Temmer}, \&
  {Gan}}]{2012ApJ...756L..41S}
{Su}, Y., {Wang}, T., {Veronig}, A., {Temmer}, M., \& {Gan}, W. 2012, \apjl,
  756, L41

\bibitem[{{Sun} {et~al.}(2015){Sun}, {Cheng}, {Ding}, {Guo}, {Priest},
  {Parnell}, {Edwards}, {Zhang}, {Chen}, \& {Fang}}]{Sun2015NatCo}
{Sun}, J.~Q., {Cheng}, X., {Ding}, M.~D., {et~al.} 2015, Nature Communications,
  6, 7598

\bibitem[{{Sun}(2013)}]{sun2013arxiv}
{Sun}, X. 2013, ArXiv e-prints, arXiv:1309.2392

\bibitem[{{Sun} {et~al.}(2013){Sun}, {Hoeksema}, {Liu}, {Aulanier}, {Su},
  {Hannah}, \& {Hock}}]{2013ApJ...778..139S}
{Sun}, X., {Hoeksema}, J.~T., {Liu}, Y., {et~al.} 2013, \apj, 778, 139

\bibitem[{{Sun} {et~al.}(2017){Sun}, {Hoeksema}, {Liu}, {Kazachenko}, \&
  {Chen}}]{Sun2017ApJ}
{Sun}, X., {Hoeksema}, J.~T., {Liu}, Y., {Kazachenko}, M., \& {Chen}, R. 2017,
  \apj, 839, 67

\bibitem[{{Tian} {et~al.}(2016){Tian}, {Xu}, {He}, \&
  {Madsen}}]{2016ApJ...824...96T}
{Tian}, H., {Xu}, Z., {He}, J., \& {Madsen}, C. 2016, \apj, 824, 96

\bibitem[{{Tian} {et~al.}(2014){Tian}, {DeLuca}, {Cranmer}, {De Pontieu},
  {Peter}, {Mart{\'{\i}}nez-Sykora}, {Golub}, {McKillop}, {Reeves}, {Miralles},
  {McCauley}, {Saar}, {Testa}, {Weber}, {Murphy}, {Lemen}, {Title}, {Boerner},
  {Hurlburt}, {Tarbell}, {Wuelser}, {Kleint}, {Kankelborg}, {Jaeggli},
  {Carlsson}, {Hansteen}, \& {McIntosh}}]{Tian2014Sci}
{Tian}, H., {DeLuca}, E.~E., {Cranmer}, S.~R., {et~al.} 2014, Science, 346,
  1255711

\bibitem[{{Tu} {et~al.}(2005){Tu}, {Zhou}, {Marsch}, {Xia}, {Zhao}, {Wang}, \&
  {Wilhelm}}]{Tu2005Sci}
{Tu}, C.-Y., {Zhou}, C., {Marsch}, E., {et~al.} 2005, Science, 308, 519

\bibitem[{{Wedemeyer} {et~al.}(2013){Wedemeyer}, {Scullion}, {Rouppe van der
  Voort}, {Bosnjak}, \& {Antolin}}]{2013ApJ...774..123W}
{Wedemeyer}, S., {Scullion}, E., {Rouppe van der Voort}, L., {Bosnjak}, A., \&
  {Antolin}, P. 2013, \apj, 774, 123

\bibitem[{{Wedemeyer-B{\"o}hm} \& {Rouppe van der
  Voort}(2009)}]{Wedemeyer2009AA}
{Wedemeyer-B{\"o}hm}, S., \& {Rouppe van der Voort}, L. 2009, \aap, 507, L9

\bibitem[{{Wedemeyer-B{\"o}hm} {et~al.}(2012){Wedemeyer-B{\"o}hm}, {Scullion},
  {Steiner}, {Rouppe van der Voort}, {de La Cruz Rodriguez}, {Fedun}, \&
  {Erd{\'e}lyi}}]{Wedemeyer2012nature}
{Wedemeyer-B{\"o}hm}, S., {Scullion}, E., {Steiner}, O., {et~al.} 2012, \nat,
  486, 505

\bibitem[{{Wiegelmann}(2007)}]{wiegelmann2007sol}
{Wiegelmann}, T. 2007, \solphys, 240, 227

\bibitem[{{Wilmot-Smith} {et~al.}(2009){Wilmot-Smith}, {Hornig}, \&
  {Pontin}}]{WilmotSmith2009ApJ}
{Wilmot-Smith}, A.~L., {Hornig}, G., \& {Pontin}, D.~I. 2009, \apj, 696, 1339

\bibitem[{{Wilmot-Smith} {et~al.}(2010){Wilmot-Smith}, {Pontin}, \&
  {Hornig}}]{WilmotSmith2010AA}
{Wilmot-Smith}, A.~L., {Pontin}, D.~I., \& {Hornig}, G. 2010, \aap, 516, A5

\bibitem[{{Wyper} {et~al.}(2016){Wyper}, {DeVore}, {Karpen}, \&
  {Lynch}}]{2016ApJ...827....4W}
{Wyper}, P.~F., {DeVore}, C.~R., {Karpen}, J.~T., \& {Lynch}, B.~J. 2016, \apj,
  827, 4

\bibitem[{{Wyper} \& {Pontin}(2014)}]{2014PhPl...21j2102W}
{Wyper}, P.~F., \& {Pontin}, D.~I. 2014, Physics of Plasmas, 21, 102102

\bibitem[{{Xue} {et~al.}(2016){Xue}, {Yan}, {Cheng}, {Yang}, {Su}, {Kliem},
  {Zhang}, {Liu}, {Bi}, {Xiang}, {Yang}, \& {Zhao}}]{Xue2016NatCo}
{Xue}, Z., {Yan}, X., {Cheng}, X., {et~al.} 2016, Nature Communications, 7,
  11837

\bibitem[{{Zhang} \& {Liu}(2011)}]{2011ApJ...741L...7Z}
{Zhang}, J., \& {Liu}, Y. 2011, \apjl, 741, L7

\end{thebibliography}
\end{document}